\documentclass[prb,twocolumn,showpacs,preprintnumbers,amsmath,amssymb]{revtex4}

\usepackage{graphicx}
\usepackage{dcolumn}
\usepackage{bm}

\begin{document}

\preprint{APS/123-QED}

\title{The influence of bond-rigidity and cluster diffusion on the self-diffusion of hard spheres with square-well interaction.}
\author{Sujin Babu}
\author{Jean Christophe Gimel}
\email{Jean-Christophe.Gimel@univ-lemans.fr}
\author{Taco Nicolai}%
\email{taco.nicolai@univ-lemans.fr}
\affiliation{Polym\`eres Collo\"{\i}des Interfaces, CNRS UMR6120,
Universit\'e du Maine, F-72085 Le Mans cedex 9, France}
\author{C. De Michele}
\affiliation{Dipartimento di Fisica,Universit\'a di Roma "La Sapienza" Piazzale Aldo Moro 2,  
00815 Roma, Italy and INFM-CRS SOFT, Universit\'a di Roma "La Sapienza" Piazzale Aldo Moro 2, 
00815 Roma, Italy}

\date{\today}
\begin{abstract}
Hard spheres interacting through a square-well potential were simulated using two different 
methods: Brownian Cluster Dynamics (BCD) and Event Driven Brownian Dynamics (EDBD). The 
structure of the equilibrium states obtained by both methods were compared and found to be almost the 
identical. Self diffusion coefficients ($D$) were determined as a function of the interaction strength. 
The same values were found using BCD or EDBD. Contrary the EDBD, BCD allows one to study the effect of 
bond rigidity and hydrodynamic interaction within the clusters. When the bonds are flexible the 
effect of attraction on $D$ is relatively weak compared to systems with rigid bonds. $D$ increases 
first with increasing attraction strength, and then decreases for stronger interaction. Introducing 
intra-cluster hydrodynamic interaction weakly increases $D$ for a given interaction strength. 
Introducing bond rigidity causes a strong decrease of $D$ which no longer shows a maximum as 
function of the attraction strength. 
\end{abstract}

\maketitle
\section{Introduction}
Suspensions of particles with attraction exhibit equilibrium states as well as non-equilibrium states like gels or glasses 
that evolve very slowly \cite{994,1030,952,730}. Numerical simulations were found to be useful for the understanding of the structure 
and the dynamics of such systems \cite{952,zacc}. The advantage of simulations is 
that large scale phenomena may be related to the microscopic trajectories of the particles. In recent years computer 
simulations have yielded valuable insight not only into equilibrium properties such as cluster size distributions and 
structure factors, but also into the evolution of the system during phase separation \cite{955,912,883} and gelation \cite{925,934,901,921,943}. 

Often Monte Carlo methods are used to study structural properties at equilibrium and molecular 
dynamics to study dynamics and kinetics \cite{1012,rapaport}. Monte-Carlo methods allow one to study relatively 
large systems and generally require less computer time to obtain equilibrium states. A main draw back of classical 
Monte Carlo methods is that the definition of time is usually unphysical so that the evolution of the system 
towards equilibrium cannot be compared to that of real systems. Molecular dynamics simulate the particle 
displacement more realistically, but the system size and time scales that can be simulated with the current 
generation of computers are still relatively small. 

The simplest model of interacting fluids is an ensemble of hard spheres that interact through a 
square well potential. Here we compare two different methods to simulate this model system: 
Brownian Cluster Dynamics (BCD) and Event Driven Brownian Dynamics (EDBD) \cite{1034}. With BCD clusters 
are constructed by forming bonds between spheres within each others interaction range with a given 
probability. With this method it is possible to account for hydrodynamic interaction within the clusters, 
though not between the clusters. It is also possible to study the influence of bond rigidity on the 
dynamics. This is important because in real systems bonds may be more or less rigid. With EDBD hydrodynamic 
interaction is ignored and the bonds are inherently completely flexible. We will show here that for reversibly 
aggregating systems bond rigidity has no influence on the structure of the steady state, but has a huge effect 
on the dynamics.  

In the following we will first describe the two simulation methods. Then we compare the structure 
factors and the cluster size distributions of homogenous equilibrium states. We will show that almost 
the same structures are obtained at steady state with both methods. The main part of paper deals with 
the self diffusion coefficient as a function of the interaction strength. We compare the results obtained 
by the two simulation methods and discuss the influence of bond flexibility and intra-cluster hydrodynamic 
interaction. 

\section{Simulation method}
We simulate hard spheres interacting through a square well potential characterized by a well depth $u$ and a well 
width $\epsilon$ using BCD and EDBD. Both simulation methods start with an ensemble of $N_{tot}$ randomly 
distributed spheres with diameter equal to unity in a box of size $L$ so that the volume fraction is defined as $\phi=(\pi/6) N_{tot}/L^3$.

\textbf{Event Driven Brownian Dynamics}. This method was described in detail in the literature \cite{1034} and we 
only resume here the principal features. Initially a random velocity is assigned to each sphere from a Gaussian distribution 
with average squared velocity $<v^2>=3kT/M$ and variance $(kT/M)^{1/2}$, where $k$ is Boltzmann's constant, $T$ is the absolute 
temperature and $M$ is the mass of the particle. Events are defined as occurrences when the sphere is at a distance unity or $1+\epsilon$ from another sphere, 
i.e. when spheres touch, or cross the interaction range. First the shortest time, $\Delta t$, before an event 
occurs is calculated. Then all spheres are moved over a distance $r=v\Delta t$, where $\Delta t$ cannot exceed a certain 
maximum value. It is important that the maximum value of $\Delta t$ is sufficiently small so that the motion is Brownian over the relevant 
length scales, i.e. the interaction range and the average distance between the nearest neighbors. 

After each sphere has been displaced the velocity of the spheres involved in the event are changed while conserving 
the energy and the momentum. When the event is a collision the spheres will bounce elastically in opposite directions. 
When the sphere enters a well its velocity is decreased because the potential energy is increased. When the sphere 
tries to leave a well it either bounces back elastically or it exits with a higher velocity. The change in the velocity 
and the probability to exit the well depend on $u$. 

The mean squared displacement of a single sphere is given by: $<r^2>=n(\Delta t)^2<v^2>$, where $n$ is the number 
of simulation steps and $\Delta t$ is equal to the maximum time step. If time is defined as $t\equiv n/(\Delta t)^2$ 
then the free diffusion coefficient of a single sphere is equal to: $D_0=kT/(2M)$. In this paper the unit of energy is 
the thermal energy. We note that in the literature often $u$ is fixed at unity and $T$ is varied.   

\textbf{Brownian Cluster Dynamics. }Spheres are considered to be in contact when they are within each others 
interaction range, i.e. when the center to center distance is smaller than $1+\epsilon$. In the so-called cluster 
formation step, spheres in contact are bound with probability $P$. Alternatively, bonds are formed with probability 
$\alpha$ and broken with probability $\beta$, so that the $P=\alpha/(\alpha + \beta)$. In the latter case one can vary 
the kinetics of the aggregation from diffusion limited ($\alpha=1$) to reaction limited ($\alpha\rightarrow 0$) 
with the same $P$ and thus the same degree of reversibility. Clusters are defined as collections of bound spheres 
and monomers are clusters of size one. After this procedure $N_c$ clusters are formed. We mention that more complex interaction potentials can be simulated by making 
$P$ a function of the distance between two spheres.   

The ratio of the number of bound ($\nu_b$) to free contacts ($\nu$-$\nu_b$) is given by the Boltzmann 
distribution: $\nu_b/(\nu-\nu_b)=\exp(\Delta H)$, where $\Delta H$ is the enthalpy difference between 
bound and free contacts. The formation of $\nu_b$ randomly distributed bonds over $\nu$ contacts leads to a decrease 
of the free energy equal to $u$ per contact. This decrease may be written as the sum of the decrease of the enthalpy 
and the gain of the entropy: $\nu \cdot u=\nu_b \Delta H-T\Delta S$. The latter is determined by the number of ways 
$\nu_b$ bonds can distributed over $\nu$ contacts: $T\Delta S=\ln(\nu!/[\nu_b!(\nu-\nu_b)!])$. Noticing that 
P=$\nu_b/\nu$, we can express $P$ in terms of $u$:
\begin{equation} 
P=1-\exp(u) 
\label{1} 
\end{equation} 

The cluster construction step is followed by one of three different movement steps that each simulates a different type of cluster dynamics. 

\textbf{BCD1}. $N_{tot}$ times a sphere is randomly selected and an attempt is made to move it a distance $s$ in a random 
direction. The movement is accepted if it does not lead to overlap with any other sphere in the system and if it does not 
lead to the separation of bound spheres beyond the interaction range. Again it is important to choose the step size $s$ 
sufficiently small so that the motion is Brownian over the relevant length scales. We have found that the results on the 
equilibrium structure were independent of the step size if $s$ was at least five times smaller than the interaction range 
and at least three times smaller than the average distance between nearest neighbors \cite{882,955}. 

The mean squared displacement of a single sphere is given by: $<r^2>=ns^2$ where $n$ is again the number 
of simulation steps. Time was defined as $t\equiv n/s^2$, so that the free diffusion coefficient of single 
spheres is equal to: $D_0=1/6$. 

\textbf{BCD2} $N_{c}$ times a cluster is randomly selected. An attempt is made to move the whole 
cluster over a distance $s/d$ in a random direction with $d$ the diameter of the cluster. By definition this cooperative 
movement never leads to bond breaking. The movement is refused if it leads to overlap of any of the spheres in the 
clusters with other spheres in the system. The free diffusion 
coefficient of single spheres is thus still $1/6$, but the free diffusion coefficient of clusters is $1/(6d)$. 

\textbf{BCD3 }This movement step is a combination of the previous two. First the movement step BCD1 is executed 
and the displacement of the centers of mass of each cluster is calculated. Then each cluster is given an additional 
displacement in the same direction so that the total displacement of the center of mass is the same as would be 
obtained by the movement step of BCD2. Again displacements are refused if they lead to overlap. As for movement 
step BCD2, the free diffusion coefficient of single spheres is $1/6$, and that of larger clusters is $1/(6d)$. 
A lower degree of flexibility can be simulated by performing movement step BCD1 with a lower frequency than movement step BCD2. 

The methods EDBD, BCD1 and BCD3 simulate systems with flexible bonds, while BCD2 simulates systems with 
rigid bonds. Using EDBD and BCD1 the effective friction coefficient of clusters is proportional to their aggregation 
number (so-called Rouse dynamics)\cite{1004}, while for BCD2 and BCD3 it is proportional to their radius (so-called Zimm dynamics)\cite{949}. 
It is, of course, straightforward to modify BCD2 to simulate systems with rigid bonds in which the friction coefficient 
is proportional to their aggregation number. This has not been done here, because in reality hydrodynamic interaction 
causes the free diffusion coefficient of clusters in solution to be inversely proportional to their radius \cite{776}. The 
movement steps of EDBD and BCD1 are similar and one expects that diffusion coefficients obtained by these methods are close.  

BCD may be considered a Monte Carlo type simulation, but if one is interested only in the structural properties 
at equilibrium it is more efficient to use other Monte-Carlo techniques that do not yield realistic kinetics or 
dynamics. BCD does not fulfill the condition of detailed balance, but does lead to a steady state independent of 
the starting configuration, which shows that it fulfills the condition of balance \cite{vasilious}. The same steady state is 
reached for each of the three movement steps with one exception: BCD2 does not lead to crystallization. The reason is that 
the pathway to form crystals is extremely unlikely when the bonds are rigid. BCD2 is therefore an excellent method to explore to 
properties of attractive spheres while avoiding crystallization \cite{955}.    

\section{Results and discussion}
\subsection{Equilibrium structure }
The strength of the attraction is 
determined by the well width and the well depth. However, the equilibrium structure obtained at different $\epsilon$ is close if 
$u$ is chosen such that the second virial coefficient is the same, especially if $\epsilon$ is small\cite{787,955}. $B_2$ may be 
written as the sum of a repulsive (excluded volume) part and an attractive part: $B_2=B_{rep}-B_{att}$, with 
\begin{eqnarray}
&&B_{rep}= 4   \nonumber \\
&&B_{att}=4\frac{P}{(1-P)}\left({(1+\epsilon)^3-1}\right) \nonumber 
\end{eqnarray}
where the unit of $B_2$ is the volume of a sphere.

As mentioned in the introduction strong attraction between the spheres leads to phase separation, while weak attraction 
leads to a homogeneous equilibrium state containing a distribution of transient clusters. We have characterized these states in terms of the static structure factor ($S(q)$) and the cluster size distribution.

\begin{figure}[ht]
\resizebox{0.45\textwidth}{!}{\includegraphics{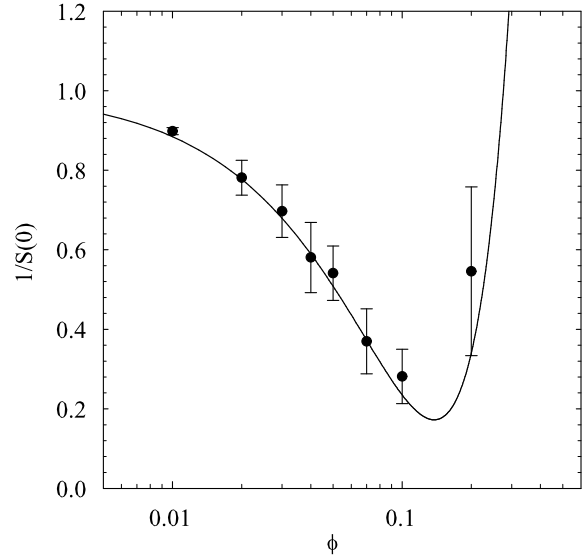}}
\caption{Concentration dependence of the compressibility for equilibrium systems at $\epsilon=0.5$ for $B_{2}=-6$ obtained by BCD. 
The solid line represents a calculation using the second and third virial coefficients, see text. The error bars 
represent the $95\%$ confidence based on the results of $8$ simulations. }
\label{f.3}
\end{figure}

The structure factor at $q\rightarrow 0$ is inversely proportional to the compressibility and can be expressed in terms of a 
virial expansion at small volume fractions: $1/S(0)=1+2B_2\phi +3B_3\phi^2$.   For hard spheres interacting with a square well 
potential the second and third virial coefficients have been calculated analytically \cite{892}. Figure(\ref{f.3}) shows a 
comparison of $1/S(0)$ as a function of $\phi$ obtained from BCD simulations with the values calculated using the virial expansion. 
There is good agreement up to $\phi=0.1$ beyond which higher order virial terms become important.   

\begin{figure}[ht]
\resizebox{0.45\textwidth}{!}{\includegraphics{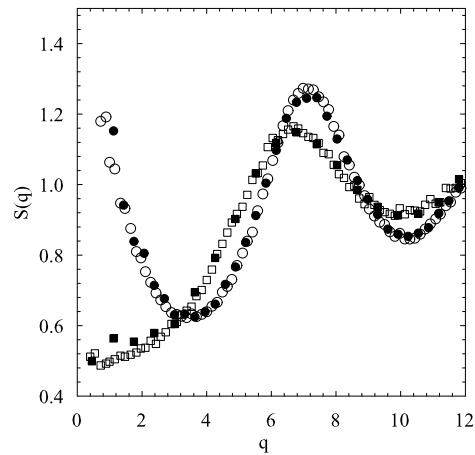}}
\caption{Comparison of the structure factor of equilibrium systems at $\phi=0.15$ and $\epsilon=0.1$ 
obtained by BCD (filled symbols) and EDBD (open symbols) with $B_{2}=2$ (squares) or $B_{2}=-2$ (circles).}
\label{f.1}
\end{figure}

Figure(\ref{f.1}) compares 
$S(q)$ obtained with BCD and EDBD at several conditions. Within the uncertainty range the same structures were observed 
with the two methods.

\begin{figure}[ht]
\resizebox{0.45\textwidth}{!}{\includegraphics{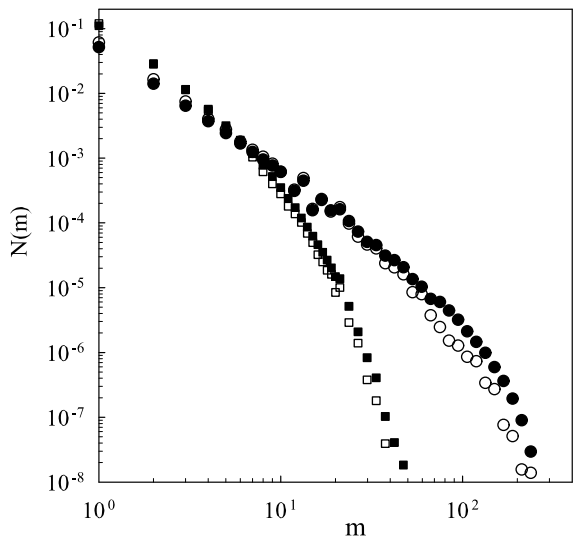}}
\caption{Comparison of the cluster size distribution of equilibrium systems at $\phi=0.15$ and $\epsilon=0.1$ obtained 
by BCD (filled symbols) and EDBD (open symbols) with $B_{2}=2$ (squares) or $B_{2}=0$ (circles).}
\label{f.2}
\end{figure}

The cluster size distribution represents a more precise characterization. Clusters of bound spheres can be 
formed by connecting spheres in contact with probability $P=1-exp(u)$ as defined above. A detailed analysis of 
these cluster size distributions has been reported elsewhere. Here, we have analysed the size distribution of clusters 
formed by connecting all contacts in order to facilitate comparison between BCD and EDBD. Figure(\ref{f.2}) shows a 
comparison of $N(m)$ i.e. the average density of clusters consisting of m spheres at $\phi=0.15$ and $\epsilon=0.1$ for 
two values of $B_2$: 2 and 0. The width of the distribution increases with increasing $u$ until at a critical value a 
system spanning transient network of spheres in contact is formed. There is a very small, but systematic difference between 
the cluster size distributions obtained with BCD and EDBD. Slightly larger clusters are formed with BCD. Nevertheless, we may conclude from these examples and similar comparisons at other conditions 
that the equilibrium structures obtained by BCD and EDBD are almost the same.  

\section{Diffusion}
\begin{figure}[ht]
\resizebox{0.45\textwidth}{!}{\includegraphics{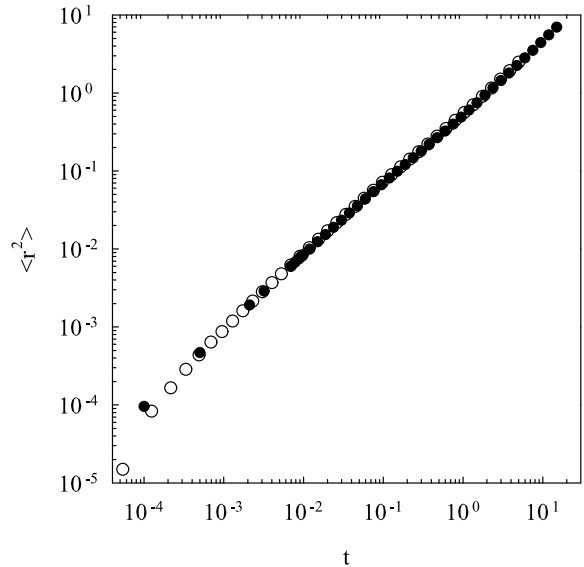}}
\caption{Comparison of the mean square displacement of non-interacting hard spheres obtained by BCD 
(filled symbols) and EDBD (open symbols).}
\label{f.4}
\end{figure}

\begin{figure}[ht]
\resizebox{0.45\textwidth}{!}{\includegraphics{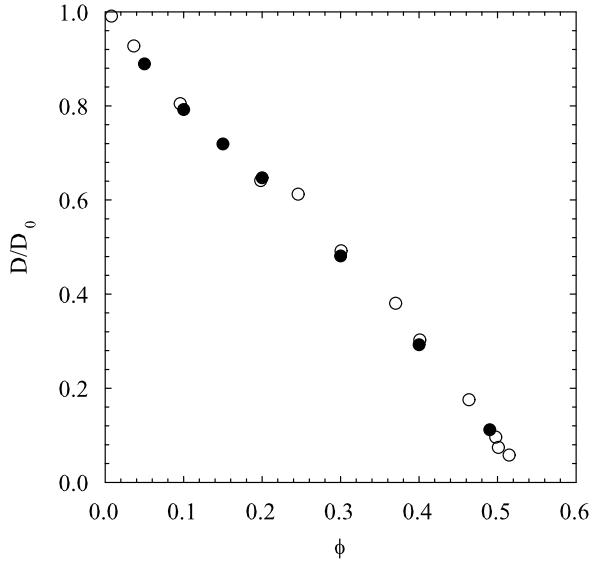}}
\caption{Comparison of the decrease of the diffusion coefficient with increasing volume fraction for 
non-interacting hard spheres obtained with BCD (filled symbols) and EDBD (open symbols)\cite{1034}.}
\label{f.5}
\end{figure}

It can be shown that random displacement with constant step size as done with BCD leads to Brownian diffusion at distances 
much larger than the step size \cite{214,1035}. For a proper comparison of the dynamic properties of BCD and EDBD simulations 
one has to ensure that the time scales and the free diffusion coefficients are the same, which can be done by choosing $\Delta t=s$ 
and $kT/M=1/3$. Figure(\ref{f.4}) shows a comparison of the average mean square displacement of hard spheres obtained from BCD and 
EDBD at $\phi$=0.3. In both simulations $<$r2$>$ becomes proportional to t and the diffusion coefficient can be calculated as 
$D=<r^2>/6t$. Figure(\ref{f.5}) shows that the same $\phi$ dependence of $D$ is obtained by the two methods within the uncertainty 
of the simulations. $D$ decreases with increasing volume fraction due to steric hindrance and the diffusion slows down 
critically at the so-called glass transition, which has been the subject of intensive investigation \cite{992,907}. 

When we introduce attraction between the spheres we need to consider cooperative cluster motion and bond flexibility. EDBD allows 
only one type of motion, but using BCD one can choose between different movement steps.  

\begin{figure}[ht]
\resizebox{0.45\textwidth}{!}{\includegraphics{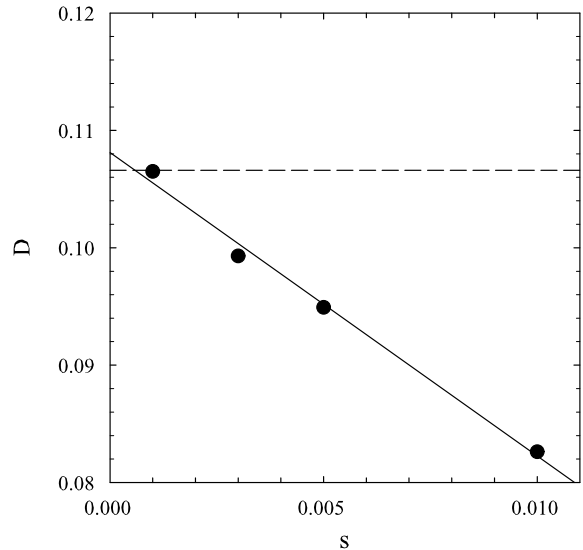}}
\caption{Dependence of the diffusion coefficient on the step size simulated using BCD1 ($\phi=0.49$, $B_{att}=6$, $\epsilon=0.1$). The dashed line 
represents the result from EDBD. We note that at $\phi=0.49$ the average distance between randomly distributed spheres is $0.014$\cite{1028,802}.}
\label{f.6}
\end{figure}

We have simulated the mean square displacement of spheres using EDBD and BCD with the three different movement steps described above. 
In each case diffusive motion was observed for large $t$ and the diffusion coefficient could be determined. BCD1 and EDBD simulate the 
same situation and therefore the results should be the same. Figure(\ref{f.6}) shows an example of the dependence of $D$ on the step 
size obtained using BCD1. It appears that the exact value of $D$ is more sensitive to the step size than the static structure 
factor or the cluster size distribution since the latter did not depend significantly on the step size for $s<0.5$. The value 
extrapolated to $s=0$ is the same as the value found with EDBD within the simulation error. Similar results were obtained 
at different volume fractions and interaction strengths. The fact that these very different simulation methods lead to the 
same results, strengthens confidence in both methods. In terms of computational efficiency both methods are equivalent.
  
\begin{figure}[ht]
\resizebox{0.5\textwidth}{!}{\includegraphics{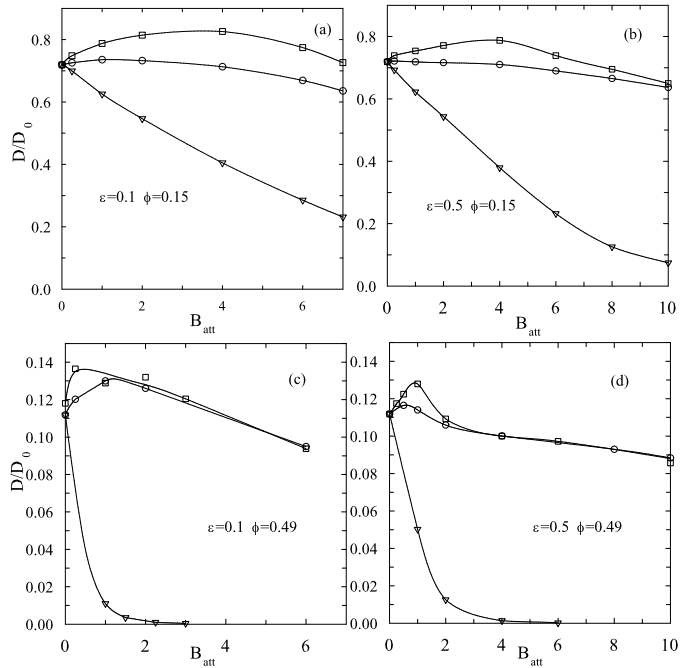}}
\caption{Variation of the diffusion coefficient with increasing attraction obtained from BCD1 (circles), 
BCD2 (triangles), and BCD3 (squares) at two different volume fractions and two different interaction widths. 
The interaction strength is expressed in terms of the attractive part of the second virial coefficient.}
\label{f.7}
\end{figure}

A comparison of the dependence of $D$ on $B_{att}$ obtained with BCD using the 3 different movement steps is 
shown in fig.(\ref{f.7}) for two different volume fractions (0.15 and 0.49) and two different well widths (0.1 and 0.5). 
The range of $B_{att}$ that can be explored is limited by the liquid-liquid or liquid-crystal phase separation that occurs 
at strong attraction. The values of $D$ shown in fig.(\ref{f.7}) were obtained at a single value of $s$, but for a few 
examples the effect of $s$ was determined, which showed that they were about $10\%$ smaller than the values 
extrapolated to $s=0$. 

$D$ decreases strongly with increasing attraction when the bonds are rigid (BCD2). In this case the displacement 
of bound spheres is equal to that of the center of mass of the clusters to which they belong. The size of the 
clusters increases rapidly with increasing attraction and beyond a critical value a transient (bond) percolating 
network is formed. Spheres that are part of the network are immobile until the bonds that tie them to the network 
are broken. A detailed study of the diffusion coefficient of square well fluids forming rigid bonds using BCD2 has been reported 
recently \cite{1038}. It was shown that $D$ decreases with increasing $B_{att}$ following a power law for large 
$B_{att}$ and only becomes zero when the bonds are irreversible, i.e. $B_{att}\rightarrow \infty$. 

In comparison, the influence of attraction on $D$ is weak when the bonds are flexible, i.e. using EDBD, BCD1 or 
BCD3. The difference between methods BCD1 (EDBD) and BCD3 is that for the latter clusters move faster 
(Zimm dynamics) so that $D$ is slightly larger. The effect increases with increasing cluster size and is expected 
to be maximal close to the percolation threshold. The values of $B_{att}$ at the bond percolation thresholds are 
about $6$ at $\phi=0.15$ for both well widths, while at $\phi=0.49$ they are 0.5 and 1.2 for $\epsilon=0.1$ 
and $0.5$, respectively. The difference between the two methods decreases for larger $B_{att}$ when most spheres 
belong to the percolating network that has no center of mass movement. The few remaining free spheres are mostly 
monomers so that the movement steps BCD1 and BCD3 become equivalent.  

Regardless of the method, the effect of attraction on $D$ is qualitatively different if the bonds are 
flexible, because in that case bound spheres can move freely within the interaction range. $D$ increases 
weakly with increasing attraction until it reaches a maximum beyond which it decreases. The relative amplitude 
of the increase is very small for the volume fractions studied here, but becomes important at higher volume 
fractions \cite{907}. It is at the origin of the so-called reentrant glass transition of interacting spheres as 
a function of the attraction strength \cite{923,931}. The influence of attraction on the critical slowing down of hard 
spheres has attracted a lot of attention in the recent past and has been investigated for square well fluids 
using EDBD simulations \cite{883}. 

The appearance of a maximum diffusion coefficient can be qualitatively understood by considering two 
opposing effects. On one hand, attraction causes clustering of particles so that more space is created in 
which monomers and clusters can diffuse freely, leading to faster diffusion of the spheres. On the other hand, 
bonds restrict the motion of spheres and the long time diffusion of bound spheres is equal to 
the center of mass diffusion of the clusters to which they belong. The restriction becomes more important as 
the average bond life time increases.

When the attraction is weak the average bond life time is still small so that the effect of restriction 
is weak and the effect of creating more free space dominates leading to an increase of $D$. With increasing 
$B_{att}$ the clusters become larger and the average bond life time increases until the effect of increasing 
restriction of the movement becomes more important than the effect of increasing free volume so that $D$ 
decreases. These features are independent of the volume fraction and the well width. The effect 
of attraction on the diffusion coefficient remains small in the single phase regime if the bonds are flexible 
atleast for $\phi<0.5$.

\section{Summary}
BCD and EDBD simulations of hard spheres interacting with a square well potential lead to steady 
states that have almost the same structure factor and the cluster size distribution. 

EDBD assumes flexible bonds and ignores hydrodynamic interaction. The values of the self diffusion 
coefficient obtained by EDBD are very close to those obtained with BCD if the same assumptions are used. 
A weak maximum of $D$ is found as a function of the interaction strength caused by the opposing effects of 
increasing free volume and increasing bond life time. 

The effect of intra-cluster hydrodynamics (Zimm dynamics) and bond rigidity can be explored with 
BCD. Introducing rigid bonds leads to a strong decrease of $D$ with increasing attraction and suppresses 
the maximum. Introducing intra-cluster hydrodynamics to the system with flexible bonds weakly increases 
$D$ at a given interaction strength. 









\end{document}